\documentclass[%
 reprint,
 amsmath,amssymb,
 aps,
]{revtex4-2}

\usepackage{float}
\usepackage{graphicx}
\usepackage{dcolumn}
\usepackage{bm}
\usepackage{hyperref}
\def\be{\begin{equation}}
\def\ee#1{\label{#1}\end{equation}}
\newcommand{\ben}{\begin{eqnarray}}
\newcommand{\een}{\end{eqnarray}}
\def\lb{\label}
\def\no{\nonumber}

\begin{document}

\preprint{APS/123-QED}

\title{Slow-roll inflation from a geometric scalar-tensor model with self-interacting potentials}

\author{Abra\~{a}o J. S. Capistrano}\email{capistrano@ufpr.br}
\affiliation{Universidade Federal do Paran\'{a}, Departmento de Engenharia e Exatas, Rua Pioneiro, 2153, Palotina, 85950-000, Paraná/PR, Brasil\\
Federal University of Latin American Integration (UNILA), Applied physics graduation program, Avenida Tarqu\'{i}nio Joslin dos Santos, 1000-Polo Universit\'{a}rio, Foz do Igua\c{c}u, 85867-670, Paran\'{a}/PR,Brasil}

\author{Gilberto M. Kremer}\email{kremer@fisica.ufpr.br}
\affiliation{Departamento de F\'isica,Universidade Federal do Paran\'a, Av. Cel. Francisco H. dos Santos, 100, Curitiba/PR, Brasil}

\date{\today}

\begin{abstract}
We consider slow-roll inflation of a model in the context of Brans-Dicke dilaton gravity. From a two self-interacting potentials $V(\phi)$, we reproduce a Starobinsky-like potential and an exponential tail potential in a form $V(\phi)\sim(1-e^{\alpha_0\phi})$, where $\alpha_0$ denotes a constant coefficient related to the Brans-Dicke parameter $\omega$. Using the observational bounds on the spectral index $n_s$ and tensor-to-scalar ratio $r$ imposed by Planck-CMB baseline data and the BICEP2/Keck collaboration with combination with Planck 2018 and the Baryonic Acoustic Oscillations~(BAO), we obtain for both models a good agreement with current observations with  $n_s = 0.960 - 0.972$ and $r<0.02$. In addition, the large values of $\omega$ suggest a possible linkage of the inflationary regime and today's solar system bounds.
\end{abstract}

\maketitle

\section{Introduction}

Today's effective description to explain the origin of large structure formation
~\cite{Starobinsky1979,Mukhanov1981,Linde1983,Hawking1982,Hawking1983,Starobinsky1982,Guth1982pi,Bardeen1983} and inhomogeneities observed in the cosmic microwave background (CMB) relies mainly on the presumption of a cosmic inflationary regime~\cite{guth81} in the early universe. Such phase consists in a rapid and exponential increase in the cosmic scale factor ensuing the big bang initial singularity. With minimal assumptions, it proposes a simple solution to the horizon besides of explaining why the universe manifests itself as homogeneous and isotropic on large scales as observed today's as inflation naturally leads to a flat universe. 

Remarkably enough, due to its simplicity, the cosmological inflation theory led to a plethora of competing models once the nature of the scalar viz. inflaton field is still the center of current debate~\cite{Giovannini1999,Giovannini1999b,Giovannini2003,cognola2008,Myrzakulov2015,Salvio2017,Salvio2019,Oikonomou2017,Agarwal2017,Keskin2018,capistrano2023,capistrano2024,Cecchini_2024}. Besides of several
proposals, inflation lacks to define a proper form for the inflaton field. Thus, it opens a large arena for inflationary models beyond or in terms of the standard phenomenological model referred as $\Lambda$CDM that associates the Cosmological Constant $\Lambda$ with cold dark matter~(CDM). Specifically, in this context, the drawback of $\Lambda$CDM paradigm with the fluid parameter $w=-1$ decays at a rapid rate at the end of the inflationary period~\cite{Castello2021}. This turns a nonviable way of relating the inflaton field as playing the role of $\Lambda$ in the inflationary regime, and makes apart the cosmological regimes of early and late universe. An extensive review of inflationary models can be found in~\cite{Martin2013}.

From the observational viewpoint, the current data imposes a stringent constraint on the main inflationary parameters namely the tensor-to-ratio $r$ which pinpoints the resulting primordial gravitational waves in inflationary period and the spectral tilt $n_s$ that defines the scale dependency of the density perturbation power spectrum. For instance, Planck collaboration  indicate that $n_s = 0.9649 \pm 0.0044$ at the $68\%$ confidence level (C.L.)~\cite{2020_CMB} and $r<0.056$ at the $95\%$ C.L.~\cite{planckinflation}. Concerning the implications for the inflationary model for single field inflation, the Starobinsky potential is highly favoured by Planck-CMB observations. The most stringent constraint comes from the BICEP/Keck collaboration setting $r < 0.036$ at a $95\%$ C.L. \cite{Ade_2021}. 

In this paper, we work with Brans-Dicke scalar-tensor theory of gravity that assumes the metric tensor $g_{\mu\nu}$ and a scalar field $\phi$~\cite{Bransdicke,bransdicke2,Tahmasebzadeh_2016} nonminimally coupled. Originally motivated by Mach's principle, the scalar field has a peculiar inner nature and is determined by the full matter content of the Universe. The application of Brans-Dicke theory of gravity to inflationary scenario was originally made in~\cite{daile} where bubble nucleation guarantees a proper phase transition of the Universe from false-- to true--vacuum phase. Such structure attracted much attention due to the fact it opened a new arena to investigate the different dynamics of the gravitational field, e.g., applications in the context of String cosmology~\cite{GASPERINI1993317,GASPERINI20031} or in the current dark energy paradigm~\cite{Riess_1998,Perlmutter_1999,DeFelice:2010jn,luongo2018}, quasi-quintessence field~\cite{D’Agostino_2022}, K-essence models~\cite{Luongo_2024}, symmetry breaking~\cite{BELFIGLIO2024101458}, etc. Moreover, a renewed interest on the theory is related to interpret the scalar field  as the inflaton field and to constrain the dimensionless  coupling constant parameter, $\omega$, which cannot be determined from theory \textit{per se}. \textit{A priori}, it requires a mechanism to understand different values it may have in different cosmological scenarios~\cite{PhysRevD.60.043501,PhysRevD.98.043516}. For instance, the standard General Relativity~(GR) theory is recovered at the limit $\omega\rightarrow \infty$. Based on time-delay experiments impose the limits $\omega > 500 \gg 1$~\cite{will} and recent large limit with at least $\omega > 1000$ considering gravitational radiation from compact binary systems~\cite{justinberti} for a massive variant of the Brans-Dicke gravity. In the present framework, we analyze the resulting values of $\omega$ in order to study the behavior of the inflationary cosmic evolution without braking today's solar system bounds.

The paper is organized in sections. The second section aims at the essentials of the present model in the context of Brans-Dicke gravity and the obtainment of the slow-roll parameters. In the third section, we present the slow-roll analysis for both considered models that resembles a Starobinsky-like model and potentials with exponential tails~\cite{Goncharov:1984jlb,Stewart95,DVALI199972,Cliff2002,Cicoli_2009,Martin2013}. The curve behavior of such potentials for different values of $\omega$ parameter are also discussed. An analysis of the results contrasted with the observational data are also implemented with Planck-CMB baseline~\cite{2020_CMB,planckinflation} and the BICEP2/Keck collaboration with combination with Planck 2018 and the  Baryonic Acoustic Oscillations~(BAO)~\cite{Ade_2021}. Finally, the conclusion and prospects are presented in the \textit{Remarks} section. 

\section{Essentials on geometrical scalar-tensor models}

The aim of this work is to determine the slow-roll parameters for an inflationary period based on the equations of a scalar-tensor theory where a scalar field is associated with a dilaton field. For the sake of completeness we summarize below the basic equations derived in~\cite{kremer2020}.

The starting point is the definition of an action $\mathcal{S}_{JF}$ on the Jordan frame
\begin{equation}
\mathcal{S}_{JF} = \int d^{4}x \sqrt{-g}e^{-\phi}\left[ R + \omega \partial^{\alpha}\phi\partial_{\alpha}\phi - e^{-\phi}V\left(\phi \right)\right],\label{Sjf}
\end{equation}
which is a function of the scalar field $\phi$ minimally coupled to gravity and its self-interacting potential $V(\phi)$. The parameter $\omega$ represents a dimensionless coupling constant and  $R$ the Ricci scalar which is calculated with the affine connection 
\ben
\Gamma^{\alpha}_{\mu\nu} = \widetilde{\Gamma}^{\alpha}_{\mu\nu} - \frac{1}{2}g^{\alpha\beta}\left( g_{\mu\beta}\partial_{\nu}\phi +  g_{\nu\beta}\partial_{\mu}\phi -  g_{\mu\nu}\partial_{\beta}\phi\right), \qquad \label{wc1}
\een
where
\ben
\widetilde{\Gamma}^{\alpha}_{\mu\nu} = \frac{1}{2}g^{\alpha\beta}\left( \partial_{\nu}g_{\mu\beta}+  \partial_{\mu}g_{\nu\beta} -  \partial_{\beta}g_{\mu\nu}\right),\label{WeylConnection}
\een
denotes the  Christoffel symbols. The relations in Eqs.(\ref{wc1}) and (\ref{WeylConnection}) result from the non-metricity condition
\ben
\nabla_{\alpha}g_{\mu\nu}= \partial_{\alpha}\phi~g_{\mu\nu}, \qquad 
\label{wc}
\een
commonly used in scalar-tensor models~\cite{Romero_2012}. Concerning notation, Greek indices runs from 1 to 4.

By definition of the  action $\mathcal{S}_{EF}$ on the Einstein frame, we write the following functional as
\begin{equation}
\mathcal{S}_{EF} = \int d^{4}x \sqrt{-\Tilde{g}}\left[ \Tilde{R} + \omega \partial^{\alpha}\phi\partial_{\alpha}\phi - V\left(\phi \right)\right].\label{Sef}
\end{equation}
The basis of the physical equivalence between Jordan and Einstein frame resides in a fact that Eqs.(\ref{wc1}) and (\ref{wc}) are invariant by Weyl transformations
\begin{equation}
\Tilde{g}^{\mu\nu}= e^{-f}g^{\mu\nu}\;,\;\Tilde{\phi}= \phi+ f,\label{weyltrans}
\end{equation}
in which $f$ is an arbitrary scalar and can be chosen, for convenience, to obtain Eq.(\ref{Sef}) from Eq.(\ref{Sjf}) by considering  that $\Tilde{R}=e^{-f}R$, since the Ricci scalar is not invariant by Weyl transformations, while the Ricci tensor is. Indeed, with $\sqrt{-\Tilde{g}}=e^{2f}\sqrt{-{g}}$ we have
\ben
&&\int d^{4}x \sqrt{-\Tilde{g}}\left[ \Tilde{R} + \omega \partial^{\alpha}\phi\partial_{\alpha}\phi - V\left(\phi \right)\right]
\\\no
&&\!=\!\!\int \!d^{4}x \sqrt{-{g}}e^{f}\big[ e^f\Tilde{R} +e^f\omega e^{-f}g^{\mu\nu}\omega \partial_{\mu}\phi\partial_{\nu}\phi - e^fV\big(\phi \big)\big].\label{Sef1}
\een

From the variational method of Palatini the metric tensor and the affine connection are independent variables, so that from the variation of the action (\ref{Sjf}) with respect to the metric tensor ${g}_{\mu\nu}$, we obtain Einstein's field equation 
\ben\lb{E1}
{R}_{\mu\nu}-\frac12 {R} g_{\mu\nu}=-{T}^\phi_{\mu\nu},
\een
where the energy-momentum tensor reads
\ben
T_{\mu\nu}^\phi=  \omega \left(\partial_{\mu}{\phi} \partial_{\nu}{\phi} -\frac{1}{2}\partial^\alpha{\phi}\partial_\alpha{\phi}g_{\mu\nu}\right) + \frac{1}{2}e^{-\phi}V {g}_{\mu\nu}. 
\een

Einstein's field equation (\ref{E1}) can be rewritten as
\ben
\Tilde{R}_{\mu\nu}-\frac12 \Tilde{R} {g}_{\mu\nu}=-\Tilde{T}_{\mu\nu}.
\een
Here $\Tilde{R}_{\mu\nu}$ and $\Tilde{R}_{\mu\nu}$ are the standard Ricci tensor and Ricci scalar calculated with Christoffel symbols $\widetilde{\Gamma}^{\alpha}_{\mu\nu}$ and the expression for the energy-momentum tensor is given by
\ben\no
    \Tilde{T}_{\mu\nu} = T^{{\phi}}_{\mu\nu}-\Tilde{\nabla}_{\mu}(\partial_{\nu}{\phi}) 
    -\frac 12 \partial_{\mu}{\phi}\partial_{\nu}{\phi}
   \\ -{g}_{\mu\nu}
    \left(\frac{1}{4}\partial_{\alpha}{\phi}\partial^{\alpha}{\phi}-\Tilde{\nabla}_{\alpha}(\partial^{\alpha}{\phi})\right). \label{kg0}
\een
Note that the covariant derivatives are related with the Christoffel symbols 
$
\Tilde{\nabla}_{\mu}(\partial_{\nu}{\phi})=\partial_\mu(\partial_\nu\phi)-\widetilde{\Gamma}^{\alpha}_{\mu\nu}\partial_\alpha\phi.$

 The variation of the action (\ref{Sjf}) with respect to the scalar field ${\phi}$, leads to the Klein-Gordon equation
\begin{equation}
   \Tilde{\Box}{\phi} - \partial^{\alpha}{\phi}\partial_{\alpha}{\phi} + \frac{e^{-\phi}}{2\omega}\frac{d V}{d \phi} = 0 .\label{KG}
\end{equation}

In a spatially flat universe, the Friedmann--Robertson--Walker (FRW) metric reads
\begin{equation}
ds^{2} = dt^{2}-a^{2}(t)\left[  dx^{2} +  dy^{2} +  dz^{2}  \right],\label{dsFW}
\end{equation}
where $a(t)\equiv a$ denotes the cosmic scale factor.

Following the development shown in~\cite{kremer2020}, an equivalent expression to Eq.(\ref{KG}) can be obtained in the Jordan frame from the Euler–Lagrange equation for the scalar field $\phi$ for a point-like Lagrangian $\mathcal{L}_{JFp}$ as given by 
\ben\no
\mathcal{L}_{JFp} = e^{-\phi}\bigg[\left(\omega-\frac 32\right)a^3\dot{\phi}^2+6\left(a^2\dot{a}\dot{\phi}-a\dot{a}^2\right)
\\
-a^3e^{-\phi}V(\phi)\bigg].\label{Sjf2}
\een

The Klein-Gordon equation in the FRW-metric follows from the variation of the Lagrangian $\mathcal{L}_{JFp}$ with respect to the scalar field $\phi$, yielding
\begin{equation}
    \ddot{\phi} + 3H\dot{\phi} - \dot{\phi}^2 + \frac{e^{-\phi}}{2\omega}\frac{d V}{d\phi} = 0, \label{KG1}
\end{equation}
in analogy with Eq. (\ref{KG}). Here the overdot refers to a time derivative and $H=\dot a/a$ denotes the Hubble parameter.

\section{Slow-roll analysis of the self-interacting potentials}
In this work, the slow-roll analysis is investigated by use Eq.(\ref{KG1}) in Jordan frame. By considering the slow-roll conditions $\ddot\phi\ll3H\dot\phi$  and $\dot\phi^2\ll V(\phi)$, Eq.(\ref{KG1}) reduces to
\ben
3 H\dot{\phi}+\frac{dV_1(\phi)}{d\phi}=0\;,\label{eq:KG2}
\een
where we have introduced the self-interacting potential $V_1(\phi)$ defined by 
\ben
V_1(\phi)=V_0+\frac{1}{2\omega}\int e^{-\phi}\frac{dV(\phi)}{d\phi}d\phi\;,\label{eq:v1}
\een
wherein $V_0$ is an integration constant. 

The potential $V_1(\phi)$ is adopted for calculation of the slow-roll analysis, defined by a test potential $V(\phi)$. The two slow-roll parameters are defined by 
\ben
\varepsilon=\frac12\left(\frac{V_1(\phi)^\prime}{V_1(\phi)}\right)^2,\qquad \eta=\frac{V_1(\phi)^{\prime\prime}}{V_1(\phi)}
\een
with the prime denoting  a differentiation with respect to the scalar field $\phi$. The reduced Planck mass $m_{pl}=1/\sqrt{8\pi G}$ was set as $1$, for simplicity, and will be restated further in the numerical analysis of the phase-space in section~\ref{section4}.

In this work we shall investigate self-interacting potentials in the slow-roll analysis considering two scenarios starting from the choice of $V(\phi)$. So far, in the construction of $V(\phi)$, we have verified that allowing mixed terms with exponential functions tend to produce more viable scenarios to inflation. It is well-known that potentials on large field models~\cite{Martin2013}, chaotic inflation~\cite{Linde1983}, or logarithm corrections~\cite{Barrow2007}, are ruled out by observations providing large values for the spectral tilt $n_s \geq 1$. Thus, we resort to two scenarios motivated by superstring models. 

In the first scenario we adopt a potential as sum of two exponential terms in a form 
\ben
V(\phi)=V_0\left(1+ A e^{c_0\phi}+ B e^{c_1\phi}\right)\;,\label{eq:pot1} 
\een
wherein we denote 
\ben\no
&&A=\frac 43 \frac{\alpha_0}{(2\alpha_0-1)(1-3\alpha_0)}, \qquad \alpha_0=\frac{3\omega - 1}{9\omega}
\\\no
&&B=-\frac 43 \frac{\alpha_0}{(\alpha_0-1)(1-3\alpha_0)}, \een and the coefficients $c_0=1-2\alpha_0$ and $c_1=1-\alpha_0$. This form of potential generalizes the original potential produced by D-branes when sypersymmetry~(SUSY) is broken at the string scale~\cite{Dudas_2012}.

By means of Eq.(\ref{eq:v1}), Eq.(\ref{eq:pot1}) generates a Starobinsky-like potential $V_{SL}$ in a form 
\ben
V_{SL}(\phi)=V_0 (1-e^{-\alpha_0\phi})^2\;.\label{eq:starob}
\een
Thus, one finds the slow-roll pairs $(\varepsilon,\eta)$ as
\ben
&&\varepsilon= \frac{2\alpha_0^2}{\left(1-e^{\alpha_0\phi}\right)^2}\;, \label{eq:eps01}
\\
&&\eta=2\alpha_0^2\frac{(2-e^{\alpha_0\phi})}{\left(1-e^{\alpha_0\phi}\right)^2}\;.\label{eq:eta01}
\een
For simplicity of notation, we maintain the $\alpha_0$ parameter in the equations.

The value of the field at the end of inflation $\phi_{end}$ is calculated by setting $\varepsilon(\phi_{end}) = 1$. Thus we obtain the real solution in form
\begin{equation}
\phi_{end}= \frac{1}{\alpha_0}\ln{(1+\sqrt{2}\alpha_0)}\;.\label{eq:phiend01}
\end{equation}
Now we are able to calculate the number of e-foldings of inflation denoted by the quantity $N$   defined as the logarithm of the scale factor $a(t)$ such as $N = N_0 + \ln a$, and $N_0$ denotes the number of the e-folds set by the current time $a_0(t_0)$. Then, $N$ can be computed as
\begin{equation}\label{eq:efolds}
\Delta N= -\int^{\phi}_{\phi_{ini}} \frac{d\phi}{2\sqrt{\varepsilon}}\;,
\end{equation}
where $\Delta N= N-N_0$ and $\phi_{ini}$ is the initial value of the inflaton field. 

Plugging Eq.(\ref{eq:eps01}) into Eq.(\ref{eq:efolds}), we get straightforwardly
\begin{equation}\label{eq:efolds02}
2\alpha_0^2 \Delta N= \alpha_0(\phi-\phi_{ini})-(e^{\alpha_0\phi}-e^{\alpha_0\phi_{ini}})\;,
\end{equation}
which can be promptly inverted to obtain the values of the field in terms of the number of e-folds as
\ben\no\label{eq:phifolds01}
&&\phi= \frac{1}{\alpha_0}\bigg(2\alpha_0^2 \Delta N+\alpha_0 \phi_{ini}-e^{\alpha_0\phi_{ini}} 
\\
&&\;- W_{-1}\big(-\exp{(2\alpha_0^2 \Delta N+\alpha_0 \phi_{ini}-e^{\alpha_0\phi_{ini}}})\big)\bigg)\;,
\een
where $W_{-1}$ is the $-1$-branch of Lambert function. Thus, taking Eq.(\ref{eq:phiend01}) and Eq.(\ref{eq:phifolds01}), we calculate the value of the field before the end of inflation $\phi_{\star}$ for $\Delta N_{\star}= N_{end}-N$ e-folds, which is given by
\begin{equation}\label{eq:phifolds02}
\phi_{\star}= \frac{1}{\alpha_0}\left(\Xi-W_{-1}(-\exp{(\Xi)})\right)\;,
\end{equation}
where  $\Xi= -2\alpha_0^2 \Delta N_{\star}+\ln{(1+\sqrt{2}\alpha_0)}-(1+\sqrt{2}\alpha_0)$.

In the second scenario we generalize the so-called exponential tail potential, originally conceived in the context of superstring models where inflation occurs at large inflaton vacuum expectation value~\cite{OBUKHOV1993214,PhysRevD.51.6847}. Thus, we define a potential in a form 
\ben
V(\phi)=V_0\left(1 + \frac B2 e^{c_0\phi}\right)\;,\label{eq:pot2} 
\een
where $\alpha_0$, $B$ and $c_0$ are the same coefficients defined in Eq. (\ref{eq:pot1}). As a result, by using Eq.(\ref{eq:v1}), Eq. (\ref{eq:pot2}) generates an exponential inflation potential $V_{SIL}$ in a form
\ben
V_{SIL}(\phi)=V_0(1-e^{-\alpha_0\phi})\;.\label{eq:starob2}
\een
Thus, one finds the slow-roll pairs $(\varepsilon,\eta)$ as
\ben
&&\varepsilon= \frac{\alpha^2}{2\left(1-e^{\alpha_0\phi}\right)^2}\;, \label{eq:eps02}
\\
&&\eta=\frac{\alpha_0^2}{1-e^{\alpha_0\phi}}\;,\label{eq:eta02}
\een
where, for simplicity, we maintain the notation with $\alpha_0$ coefficient that carries the Brans-Dicke $\omega$ coupling parameter.

\begin{figure}
  \centering
   {\includegraphics[angle=0,width=0.48\textwidth]{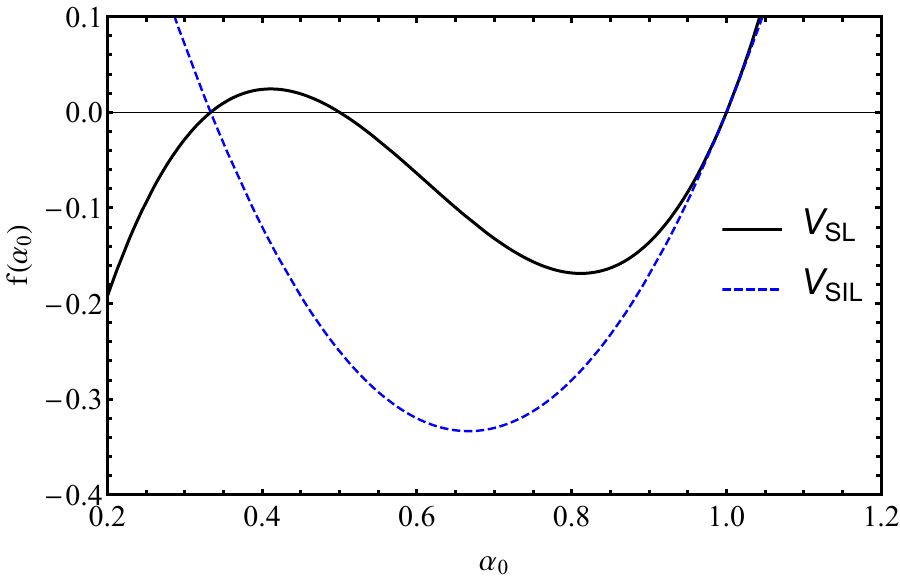}}
   \caption{Plot of $f(\alpha_0)$ the denominator  in Eqs.(\ref{eq:pot1}) and (\ref{eq:pot2}) versus $\alpha_0$. The curves represent the behavior of the potentials in Eq.(\ref{eq:pot1})~(thick black line)  and Eq. (\ref{eq:pot2})~(dashed blue line). For interpretation of the references to color in this figure legend, the reader is referred to the web version of this article.}
    \label{fig:root}
\end{figure}

The value of the field at the end of inflation $\phi_{end}$ through the condition $\varepsilon(\phi_{end}) = 1$ is given by
\begin{equation}
\phi_{end}= \frac{1}{\alpha_0}\ln{\frac{1}{2}(2+\sqrt{2}\alpha_0)}\;.\label{eq:phiend02}
\end{equation}

Plugging Eq.(\ref{eq:eps02}) into Eq.(\ref{eq:efolds}), we have the similar forms as of Eqs.(\ref{eq:efolds02}) and (\ref{eq:phifolds01}) of the previous first scenario. Thus, we obtain the slow-roll trajectory as
\begin{equation}\label{eq:efolds03}
\alpha_0^2 \Delta N= \alpha_0(\phi-\phi_{ini})-(e^{\alpha_0\phi}-e^{\alpha_0\phi_{ini}})\;.
\end{equation}
The inverse function of Eq.(\ref{eq:efolds03}) gives the values of the field in terms of the number of e-folds as
\ben\no\label{eq:phifolds03}
&&\phi= \frac{1}{\alpha_0}\bigg(\alpha_0^2 \Delta N+\alpha_0 \phi_{ini}-e^{\alpha_0\phi_{ini}} 
\\&&\qquad- W_{-1}\big(-\exp{(\alpha_0^2 \Delta N+\alpha_0 \phi_{ini}-e^{\alpha_0\phi_{ini}}})\big)\bigg),
\een
where $W_{-1}$ is the $-1$-branch of Lambert function. As a result, the value of the field before the end of inflation $\phi_{\star}$ for the potential in Eq.(\ref{eq:starob2}), is given by
\begin{equation}\label{eq:phifolds04}
\phi_{\star}= \frac{1}{\alpha_0}\left(\chi-W_{-1}(-\exp{\chi})\right)\;,
\end{equation}
where  $\chi= -\alpha_0^2 \Delta N_{\star}+\ln{\frac{1}{2}(2+\sqrt{2}\alpha_0)}-\frac{1}{2}(2+\sqrt{2}\alpha_0)$ and $\Delta N_{\star}= N_{end}-N$ e-folds.

In order to get some initial insight on the possible values of $\alpha_0$ (actually, for some values of $\omega$), we pinpoint the limiting cases where the potentials blow up, namely when the denominator generically represented by $f(\alpha_0)$ goes to zero in Eqs.(\ref{eq:pot1}) and (\ref{eq:pot2}). Thus, one obtains the roots of $f(\alpha_0)=0$ as shown in Fig.(\ref{fig:root}). For the potential defined in Eq.(\ref{eq:pot1}), it has three real roots as shown in the thick black line as $\alpha_0=\{1/3,1/2,1\}$, or, respectively, $\omega=\{+\infty, -2/3,-1/6\}$. For the potential defined in Eq. (\ref{eq:pot2}), it has two real roots as shown in the dashed blue line as $\alpha_0=\{1/3,1\}$, or, respectively, $\omega=\{+\infty, -1/6\}$. Then, we have an interesting range to explore the values of $\omega$. Note that the  Brans-Dicke theory goes over to the Einstein theory in the limit $\omega\rightarrow +\infty$. 

\section{The impact of $\omega$ on the solutions}\label{section4}

It is worth noting a discussion on the influence of the $\omega$ parameter in the solutions. 
The limits based on time-delay experiments
require a large $\omega$, i.e. $\omega > 500$ and such value should be also in conformity with inflation conditions. In the following, we initially relax the values of $\omega$ for a general analysis in order to converge to a reasonable description of inflation in this context.

\begin{figure}
  \centering
   {\includegraphics[angle=0,width=0.45\textwidth]{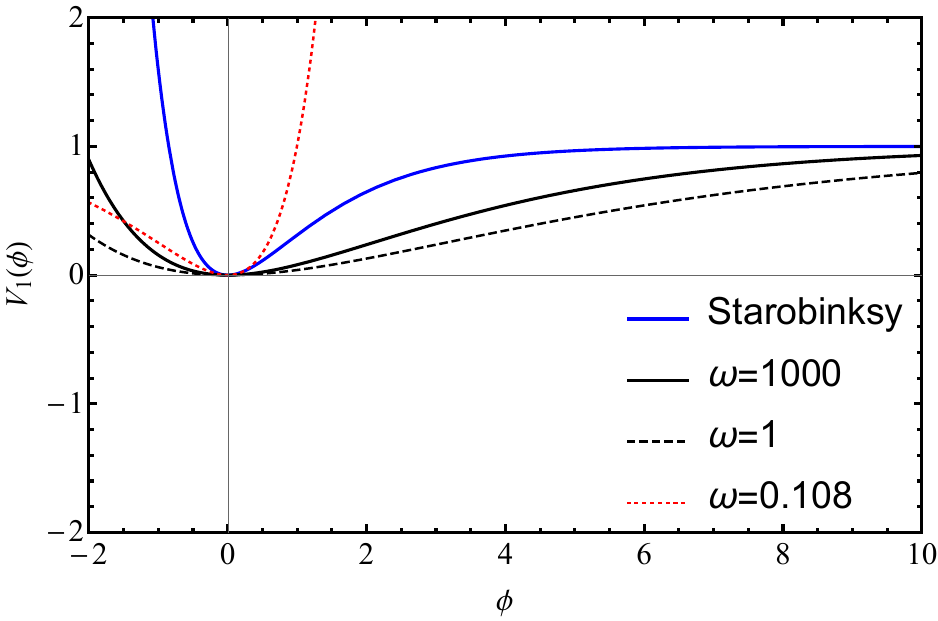}}
    {\includegraphics[angle=0,width=0.45\textwidth]{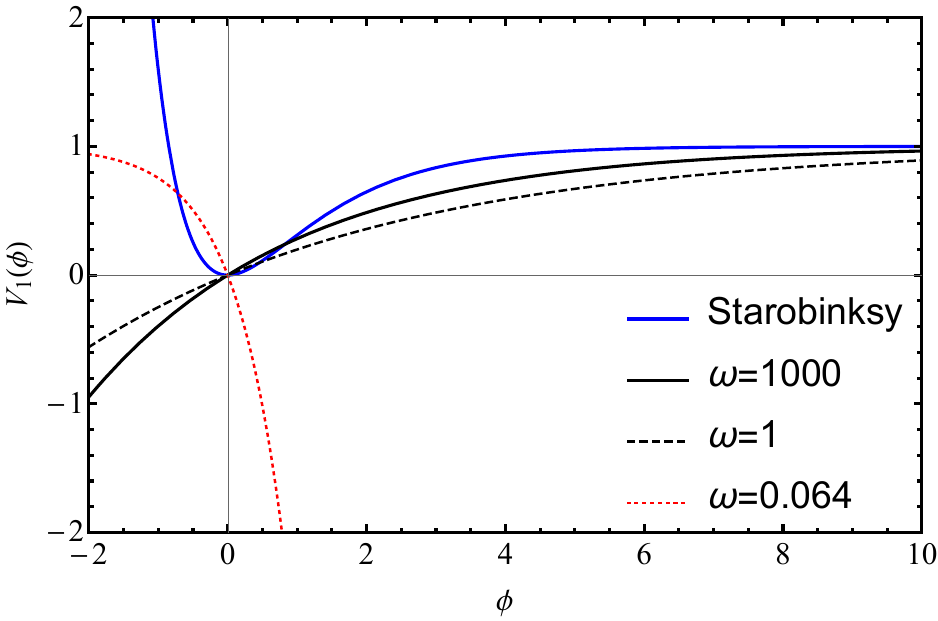}}
   \caption{Comparison between different values of the $\omega$ parameter for curves from the potentials defined in Eqs.(\ref{eq:starob}) and (\ref{eq:starob2}), as shown in the upper and lower panels, respectively. The curve from Starobinsky potential is shown by the blue thick line. For interpretation of the references to color in this figure legend, the reader is referred to the web version of this article.}
    \label{fig:comparison}
\end{figure}

As a reference, we analyze the value of the field in the end of inflation $\phi_{end}$ for the total number of e-folds, $N$, in the range $50-60$, where the field decays and goes around its minimum. For the potential in Eq.(\ref{eq:starob}), we have identified a variation of $5.78\%$ between the values of $\omega=1$ and $\omega=1000$ with $\phi_{end}\sim 1.229$ and $\phi_{end}\sim 1.158$, respectively. This percentage does not significantly change for higher values of $\omega>10$. In this regard, lower values of $\omega$ produce higher values of the field at the end of inflation until the limit $\omega\sim 0.108$ with $\phi_{end}\sim 8.31$. Below that limit, no real solutions are obtained. A drawback to adopt such lower value of $\omega$ is the fact that it alters the behavior of the potential that tends to faster exponential growth as shown in the left  panel of Figure (\ref{fig:comparison}). To avoid such situation, one needs to set as limit $\omega\sim 0.4$. 

For the exponential tail-like potential in Eq.(\ref{eq:starob2}), it does not produce high values for $\phi_{end}$, i.e., $\phi_{end}<4$. For instance, $\omega=1000$, we have $\phi_{end}\sim 0.635$, whereas for $\omega=1$ we have $\phi_{end}\sim 0.657$ that represents an increase of $3.35\%$. It maintains the same conclusion as the previous case: lower values of $\omega$ produce higher values of $\phi_{end}$. Below the limit $\omega\sim 0.064$ that gives $\phi_{end}\sim 3.434$, no real solutions are obtained. As in the case before, we have a drawback if we work with this lower limit of $\omega$ that leads to a faster exponential decay turning negative the sign of the potential as shown in the right panel of Figure (\ref{fig:comparison}). This situation can be avoided setting as limit $\omega\sim 0.4$, just as in the first case. In terms of comparison, the Starobinksy potential gives $\phi_{end}\sim 0.94$. In Figure (\ref{fig:comparison}), we also have a comparison with Starobinksy potential for different values of $\omega$. Actually, all the behavior discussed so far is in fact expected in Brans-Dicke models, i.e., the larger the value of $\omega$, the smaller the effects of the scalar field which is somewhat ``fine--tuned'' by the magnitude of $\omega$.

\begin{figure}
  \centering
   {\includegraphics[angle=0,width=0.48\textwidth]{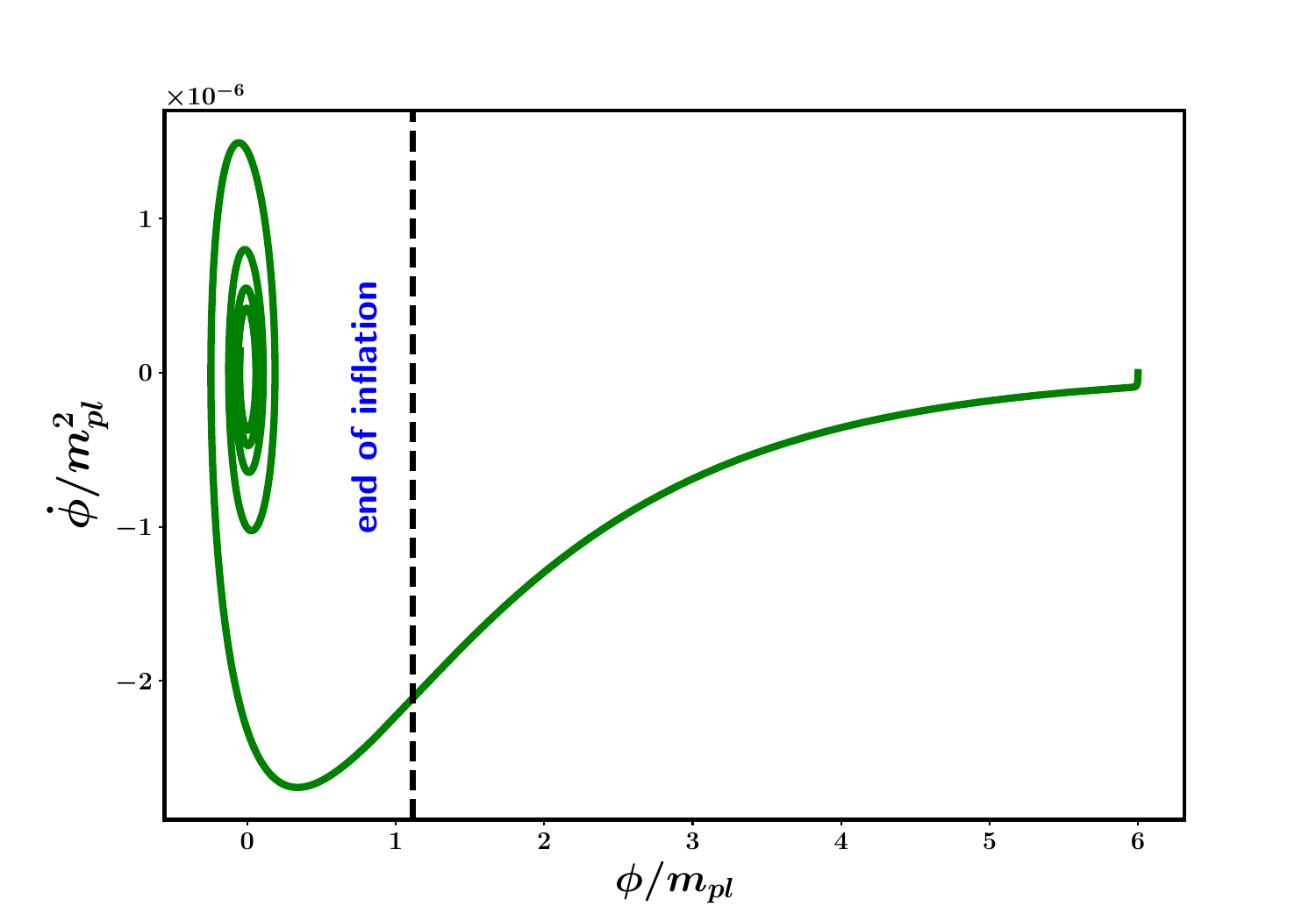}}
   \caption{The phase-space portrait $\{\phi,\dot{\phi}\}$ for the potential in Eq. (\ref{eq:starob}). A rapid convergence is achieved for the $i$-trajectories for each initial conditions $\{\phi_i,\dot{\phi_i}\}$. After the end of inflation, the oscillation of the potential is around its minimum. For interpretation of the references to color in this figure legend, the reader is referred to the web version of this article.}
    \label{fig:numdy}
\end{figure}

In order to analyze the initial conditions correlated to an adequate inflation, we perform the phase-space portrait $\{\phi,\dot{\phi}\}$ for the potential in Eq. (\ref{eq:starob})
that is represented in figure~(\ref{fig:numdy}). To perform the calculation of the dynamical system, we use the \texttt{Python} code publicly available~\footnote{See Arxiv~\url{https://doi.org/10.48550/arXiv.2212.00529} and the code at \url{https://github.com/bhattsiddharth/NumDynInflation}}. The dynamical dimensionless variables are then defined as 
$\{x = \phi/m_{pl}, y= dx/dT, A= a~m_{pl},z= H/(S m_{pl})\}$
which are the dimensionless field, field velocity, scale factor and Hubble parameter, respectively. The potential should be re-scaled as $\frac{V(\phi)}{m_{pl}^4}\rightarrow \frac{V_0}{m_{pl}^4}f(x)$, wherein $m_{pl}=1/\sqrt{8\pi G}$ is the reduced Planck mass and $f(x)$ depends on the form of the adopted potential. The quantity $T$ is the dimensionless cosmic time $T = t\;m_{pl}S$ where $t$ is the physical cosmic time and the factor $S$ to re-scale the time variable is fixed as $S= 5\times 10^{-5}$. Then, we define the initial conditions fixing $\{x_i = 6,y_i= 0, z_i = \sqrt(\frac{y_i^2}{6} + \frac{V_0~f(x_i)}{3S^2}), A_i = 2\times 10^{-3}\}$ and $V_0 = 0.5\times 10^{-10}$ for $\omega=1000$. Figure~(\ref{fig:numdy}) shows a rapid convergence of the $i$-trajectories for each initial conditions $\{\phi_i,\dot{\phi_i}\}$ and the oscillation of the potential is around its minimum after the end of inflation. 

\begin{figure}
  \centering
   {\includegraphics[angle=0,width=0.45\textwidth]{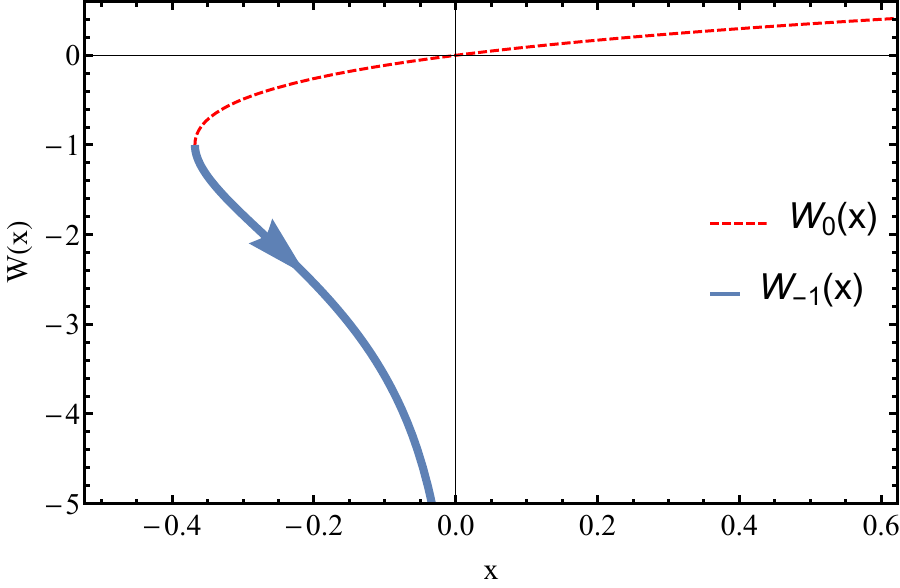}}
   \caption{Orientation of the inflation for an exponential tail-like potential of Eq.(\ref{eq:starob2}) as shown by the arrow direction in the $-1$-branch~$W_{-1}$ of Lambert function. The dashed red line denotes the trajectory in the $0$--branch $W_{0}$ of Lambert function. For interpretation of the references to color in this figure legend, the reader is referred to the web version of this article.}
    \label{fig:lambert}
\end{figure}

On the other hand, the potential in Eq.(\ref{eq:starob2}) does not oscillate around its minimum after the end of inflation. In this case, since the portrait $\{\phi,\dot{\phi}\}$ is inefficient we analyze the behavior of the argument of Lambert function. Thus, the inflation takes the orientation of the $-1$-branch of Lambert function as shown in Fig.(\ref{fig:lambert}) where $x=\phi/m_{pl}$ is given by Eq.(\ref{eq:phifolds03}) due to the fact that the term $-e^{\alpha_0\phi}$ in the argument of Lambert function is smaller than $-1$ at $\Delta N=0$ and approaches $-1$ when $x=0$. Then, the inflation proceeds at the point $-1$ between the $-1$-branch and the $0$-branch of Lambert function denoted by the red dashed line. A similar plot can be obtained for potential of Eq.(\ref{eq:starob2}) where the inflation follows the opposite direction of the trajectory shown in Fig.(\ref{fig:lambert}).

\section{Initial contrast with observational data}
\begin{figure}
  \centering
  \includegraphics[width=3.3in, height=3.2in]{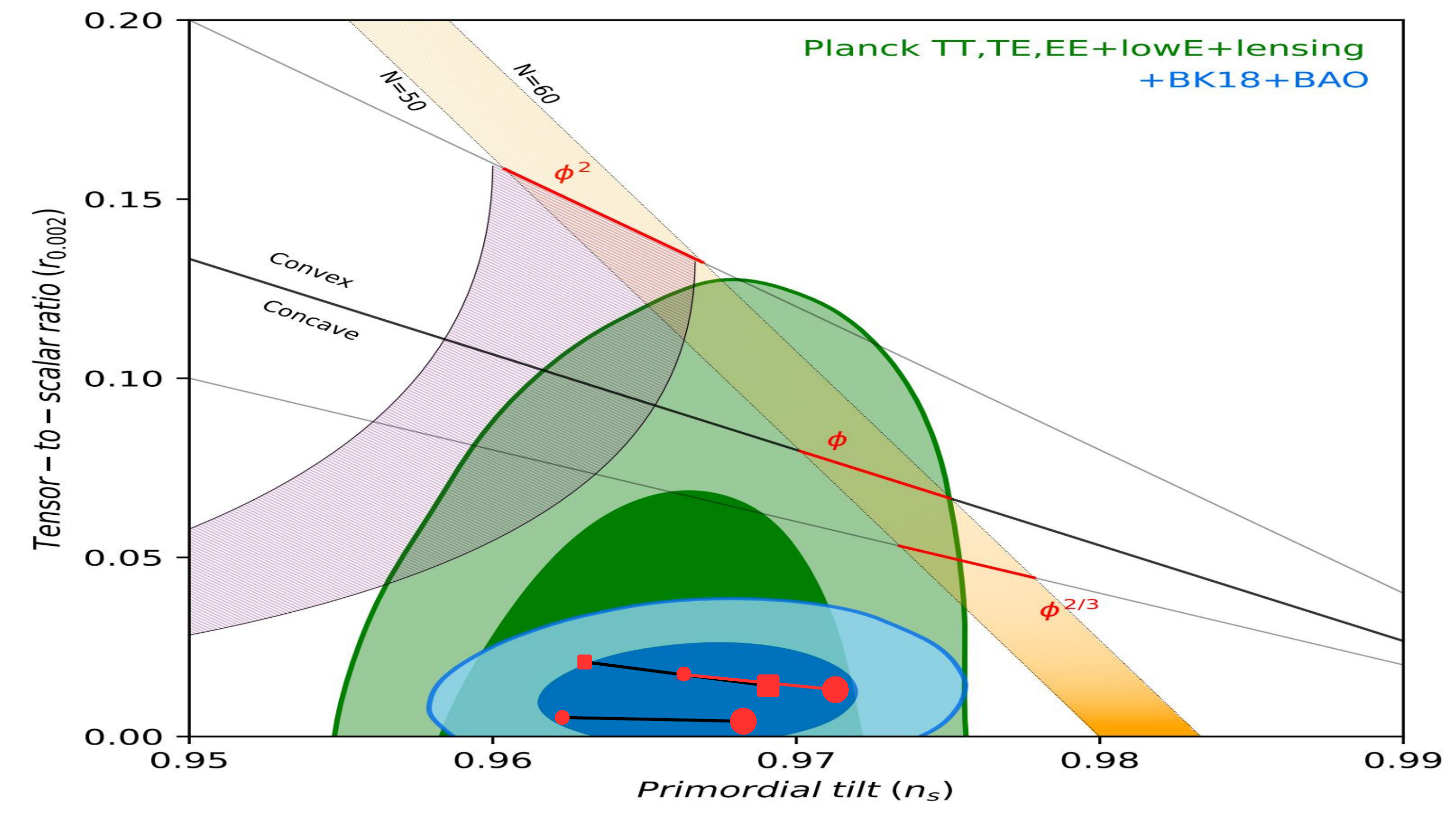}
\caption{Contours in the $r-n_s$ plane for $68\%$ and $95\%$ C.L. with the combined analyses of Planck 2018 TT,TE,EE+LowE+Lensing (green contour) and Planck 2018 TT,TE,EE+LowE+Lensing+BK18+BAO (blue contour). The monomial potential lines are shown for $N=50$ and $N=60$ e-folds. The interval of number of e-folds $N=50$ (small red square and spheres) and $N=60$~(big red square and spheres) are shown for Eqs.(\ref{eq:starob})(black line) and (\ref{eq:starob2})(red line) and compared with Starobinsky potential (black line). For interpretation of the references to color in this figure legend, the reader is referred to the web version of this article.}
\label{fig:planckBK}
\end{figure}

In order to submit the model to the scrutiny of observational data, the slow-roll parameters $(\varepsilon, \eta)$ are re-expressed as 
\begin{eqnarray}
\label{eq:tilt}
&&n_s=1-6\varepsilon+2\eta\;,\\
&&r=16\varepsilon\;,\label{eq:tensor}
\end{eqnarray}
where $n_s$ is the spectral tilt and $r$ is the tensor-to-scalar ratio. In this regard, in the following application we have fixed $\omega=1000$ for both potentials of Eqs.(\ref{eq:starob}) and (\ref{eq:starob2}). Using Eqs.(\ref{eq:eps01}) and (\ref{eq:eta01}) for potential in Eq.(\ref{eq:starob}) and Eqs.(\ref{eq:eps02}) and (\ref{eq:eta02}) for potential in Eq.(\ref{eq:starob2}), we obtain the pattern shown in Figure (\ref{fig:planckBK}).

Figure (\ref{fig:planckBK}) shows the contour plots for $68\%$ and $95\%$ C.L. from combinations of Planck 2018 baseline TT, TE, EE+LowE+Lensing (green contour)~\cite{2020_CMB} and a joint fitting with data combination of Planck 2018 TT, TE, EE+LowE+Lensing+BK18+BAO (blue contour)~\cite{Ade_2021} for selected potentials in the number of e-folds in the end of inflation for $N=50$ and $N=60$ e-folds on the pivot scale $k = 0.002$Mpc$^{-1}$. The purple region refers to natural inflation~\cite{freese90,adams93} and also monomial potentials $(\phi,\phi^{2/3})$~\cite{Lucchin1984,Martin2013} are represented. We pictorially represent the behavior of the potentials of Eqs.(\ref{eq:starob}) and (\ref{eq:starob2}) in the range $N=50-60$ e-folds by square and sphere symbols. The smaller symbols represent $N=50$ e-folds whereas the bigger ones  indicate the points calculated at $N=60$ e-folds. The values of the potential of Eq.(\ref{eq:starob}) are denoted by the upper black line with squares in the extrema end points. Moreover, the values of the potential of Eq.(\ref{eq:starob2})  are denoted by the upper red line with spheres as end points. In particular, this potential tends to a higher values of $n_s$ as compared with the other potentials. The lower black line with red sphere end points represents the Starobinsky potential. As it is shown, they fit well in the most constraining data from Planck 2018 TT, TE, EE+LowE+Lensing+BK18+BAO at $68\%$ C.L. It is worth mentioning that for both potentials, smaller values of $\omega<1$ until their lower limit, do not do produce a relevant difference with the values of $n_s$ well within the interval $n_s=0.960-0.972$ that fairly lies in the $95\%$ C.L. and $r<0.02$. On the other hand, as previously mentioned, at the lower bound limit of $\omega$, we have a fast growing or decaying potential as pointed out in Figure (\ref{fig:comparison}) which is incompatible with inflation that strongly requires a grateful exit. In both cases, such drawback does not happen for $\omega\geq 0.4$. In the range $2<\omega<3$, it also shows compatibility with anisotropic inflation in Brans-Dicke gravity~\cite{PhysRevD.98.043516}. Interestingly, higher values of $\omega$ do not generate considerable differences in the plane $r-n_s$. As presented, we have fixed $\omega=1000$ in our analysis, but if one adopts a large value, e.g., $\omega > 10^{5}$, it does not compromise the present results for the plane $r-n_s$ which is in concordance with the limits from the Cassini probe with $\omega >40000$~\cite{bertotti} and gravitational radiation from compact
binary systems~\cite{justinberti}. Remarkably, this value is the Solar System constraint for $\omega$ which suggests a natural linkage of inflationary regime with subsequent cosmology obeying today's bound~\cite{Tahmasebzadeh_2016}.

\section{Final remarks}
In this work, we investigate the slow-roll inflation in the context of a Brans-Dicke geometric scalar-tensor theory. In the Jordan frame, we obtain the related point-like Lagrangian for a FLRW metric and the resulting Klein-Gordon equation. We compute inflationary slow-roll parameters for different selected potentials reproducing a Starobinsky-like potential and potential with exponential tails to obtain the relations between ratio-to-tensor $r$ and spectral tilt $n_s$ confronted with current data of Planck-CMB baseline and the BICEP2/Keck 2018. As a result, we obtain a full agreement with current observations well within a $68\%$ C.L. Interestingly, 
high values of the Brans-Dicke parameter $\omega$ does not compromise the model in the validity range in the plane $r-n_s$ with $n_s=0.960-0.972$ and small $r<0.02$. As a primary analysis, these results suggest that the same value of $\omega$ applies to inflationary regime and to the subsequent cosmology obeying current solar system bounds $\omega \geq 10^{5}$. This gave us an optimistic scenario for further studies on late cosmic acceleration. As a subject of a future work, we will take into account the perturbation equations of the model submitting it into the scrutiny of current observational data.

\begin{acknowledgements}
A.J.S.C. acknowledges prof. Sergio Costa Ulhoa for fruitful discussions on Scalar-tensor theories. A.J.S.C. acknowledges Conselho Nacional de Desenvolvimento Cient\'{i}fico e Tecnologico (CNPq) for the partial financial support for this work (Grant No. 305881/2022-1) and Fundação da Universidade Federal do Paraná (FUNPAR, Paraná Federal University Foundation) through public notice 04/2023-Pesquisa/PRPPG/UFPR for the partial financial support (Process No. 23075.019406/2023-92);  G. M. K. acknowledges the CNPq for  financial support  (Grant No. 304054/ 2019--4).
\end{acknowledgements}

\bibliography{apssamp}

\end{document}